\documentclass[superscriptaddress,notitlepage]{revtex4-1}

\usepackage{graphicx}
\usepackage{amsmath}
\usepackage{amsfonts}
\usepackage{amssymb}
\usepackage[normalem]{ulem}
\usepackage[svgnames]{xcolor}
\usepackage{bbm}
\usepackage{bm}

\newcommand{\ten}[1]{ \mathfrak{#1} }

\renewcommand{\vec}[1]{ \mathbf{#1} }
\newcommand{\strain}{u}

\begin{document}
    \title{Multi-scale approach for strain-engineering of phosphorene}

    \author{Daniel Midtvedt}
    \email{midtvedt@chalmers.se}
    \affiliation{Department of Physics, Chalmers University of Technology, Gothenburg, Sweden}

    \author{Caio H. Lewenkopf}
    \affiliation{Instituto de F\'{i}sica, Universidade Federal Fluminense, Niter\'{o}i, Brazil}

    \author{Alexander Croy}
    \email{alexander.croy@nano.tu-dresden.de}
    \affiliation{Institute for Materials Science and Max Bergmann Center of Biomaterials,
       TU Dresden, 01062 Dresden, Germany}


    \begin{abstract}
        A multi-scale approach for the theoretical description of deformed phosphorene is presented.
        This approach combines a valence-force model to relate macroscopic strain to microscopic displacements
        of atoms and a tight-binding model with distance-dependent hopping parameters to obtain
        electronic properties. The resulting self-consistent electromechanical model is suitable for large-scale modeling of phosphorene devices.
        We demonstrate this for the case of inhomogeneously deformed phosphorene drum, which may be used as an exciton funnel.
    \end{abstract}

    \maketitle

\section{Introduction}

Atomically thin semiconductors, like molybdenum disulfide and phosphorene, have recently attracted a lot of interest 
due to their potential for optoelectronic applications\cite{miau+14,chja14,liwa+15}. 
In particular, phosphorene - a single layer of black phosphorous - has been in the focus of interest due to its anisotropic 
elastic and electronic properties\cite{xiwa+14,feya14a,qiko+14,wepe14,waku+15,wajo+15}.
Moreover, the ability to engineer the physical properties of this material by applying strain offers a range of new 
possibilities\cite{Roldan2015}. 
One example in this regard is the exciton funnel effect\cite{feqi+12,sapa+16}, 
which allows for steering excitons in a desired direction by means of an inhomogeneous band-gap. Exploiting this 
principle may offer a route towards the development of more efficient solar cells. 

The theoretical description of the elastic and electronic behavior of realistic phosphorene-based devices is, in general, 
computationally far too demanding for {\it ab initio} calculations. Valence-force models (VFM) and electronic 
tight-binding models (TB) provide viable semi-empirical alternatives. A consistent description of the influence of 
deformations on optoelectronic properties is naturally given by using the VFM to find energetically optimal positions 
of the atoms in combination with a TB model with distance-dependent hopping parameters. Within this approach 
electronic transport or optical response can be calculated retaining an accurate description of the system 
at the atomistic scale. 

On the other hand, there are many situations of practical interest where the distribution of (macroscopic) strains 
is known and one would like to use this information to infer the electronic properties of the system\cite{Roldan2015}.
For a given quantity of interest, like the electronic band-gap, one can fit the corresponding strain-induced modification using
experimental data or {\it ab initio} calculations\cite{vowa+15}. 
A more comprehensive description of the structural and electronic properties, for example in the case of non-uniform 
deformations, requires a relation between microscopic displacements of the atoms and 
the macroscopic strain. Often the Cauchy-Born rule, which states that the atomic positions within the 
crystal lattice follow the overall strain of the medium, is used\cite{Roldan2015}. However, this approximation applies only to Bravais 
lattices with a monoatomic basis\cite{er08,mile+16} and thus fails for phosphorene. Instead, this work is based 
on the strain-displacement relations obtained by minimizing the VFM of a strained unit-cell\cite{mile+16}, 
which generalize the standard Cauchy-Born approximation.

In this article we develop a theoretical framework that consistently treats both the elastic and electronic 
degrees of freedom of phosphorene \cite{micr16a,ruka14,ruyu+15} and is suitable for large-scale 
modeling of phosphorene devices.
To this end, we put forward a minimal TB model with that accurately describes the electronic band-structure of 
phosphorene in the vicinity of the $\Gamma$-point. Introducing distance-dependent hopping parameters 
and using strain-displacement relations obtained from the VFM, we derive analytical expressions that relate
the renormalization of the hopping parameters with the modifications of the band gap and effective masses caused
by strain. 
(For moderate values of strain we find that the band-gap correction obtained from the Cauchy-Born 
approximation differs by a factor $3$ from our results.)

As an application, we study the mechanical and electronic properties of a phosphorene drumhead 
subjected to uniform pressure for both the small and large deformation regimes. For different central
deflections values (the ratio between the maximal deformation height and the drum radius) 
we determine the non-uniform strain both analytically using the elasticity theory and numerically from
the VFM. The agreement is remarkable. We calculate the local band-gap an find that it depends very strongly on the 
deformation profile, and can be easily enhanced by $10$ up to $15$\%. 

The paper is structured as follows. In Sec.\ \ref{sec:model} we briefly present the lattice structure of phosphorene, 
introduce a TB model that accurately describes the low energy band structure of phosphorene, and discuss 
how to account for strain effects on the electronic properties. Sec.\ \ref{sec:results} begins with the study of the 
uniform strain case, which allows it to determine the parameters of the TB model. Next, we analyze the situation of 
non-uniform strain arising in a pressurized phosphorene drum. Finally, in Sec.\ \ref{sec:conclusions} we present our 
conclusions and outlook.


\section{Model}
\label{sec:model}

\begin{figure}[t]
    \includegraphics[width=0.33\textwidth]{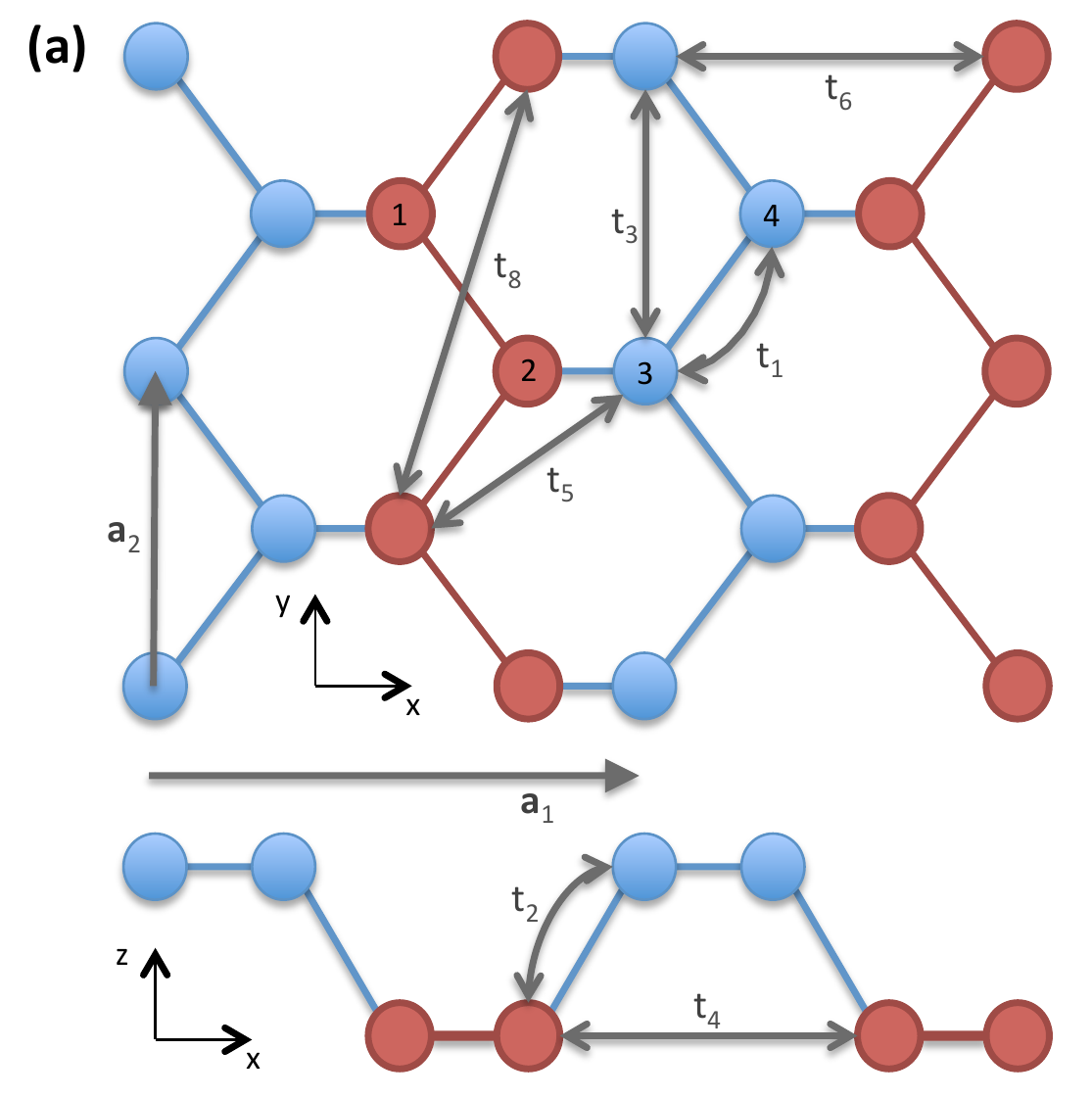}\hspace{0.1\textwidth}
    \includegraphics[width=0.33\textwidth]{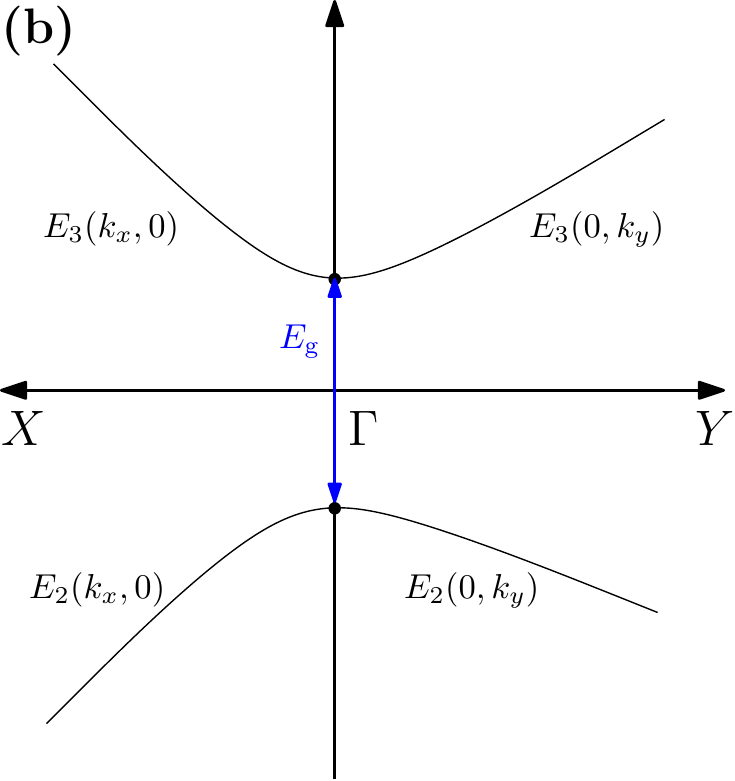}
    \caption{\label{fig:struc} 
       a) Sketch of phosphorene lattice structure indicating the hopping parameters $t_i$ used in the TB model. 
       The unit cell is spanned by the vectors $\vec{a}_1$ and $\vec{a}_2$ and consists of the four atoms labeled $1, \ldots, 4$. The armchair and zigzag directions are 
        parallel to the $x$- and $y$-axis, respectively.
        b) Sketch of the conduction ($E_3$) and valence ($E_2$) bands of phosphorene. In the armchair direction ($\Gamma-X$) 
        the band curvatures are approximately equal, while in the zigzag direction ($\Gamma-Y$) the curvature of the conduction 
        band is $4-6$ times larger than the one of the valence band.}
\end{figure}

The crystal structure of phosphorene consists of an orthorhombic lattice with lattice vectors 
$\vec{a}_1$ and $\vec{a}_2$ and four basis atoms arranged in a puckered structure, as depicted in 
Fig.\ \ref{fig:struc}(a). 
We denote the atomic positions by $\vec{R}_i$ with subindex $i$ running from $1$ to $N$, where 
$N$ is the number of atoms in the phosphorene sheet. The interatomic bond vectors are 
$\vec{b}_{ij}=\vec{R}_{j}-\vec{R}_i$ and the angle between $\vec{b}_{ij}$ and $\vec{b}_{ik}$ is denoted by 
$\theta_{jik}$. The equilibrium structure is characterized by the interatomic spacing $d\approx 2.22{\rm \AA}$ 
and intra- and inter-pucker angles $\theta_1\approx 96.5^{\circ}$ and $\theta_2 \approx 101.9^{\circ}$ \cite{kaka+86}. 
The primitive lattice vectors are thus $\vec{a}_1=2d\left( \cos(\theta_1/2)-\cos(\theta_2)/\cos(\theta_1/2),0,0\right)$ 
and $\vec{a}_2=2d\left(0,\sin(\theta_1/2),0\right)$, whereas the equilibrium bond vectors are 
$\vec{b}_{12} = d\left( \cos(\theta_1/2),  -\sin(\theta_1/2), 0 \right)$, 
$\vec{b}_{23} = d\left(-\cos(\theta_2)/\cos(\theta_1/2), 0, \sqrt{1- [\cos(\theta_2)/\cos(\theta_1/2)]^2} \right)$
and
$\vec{b}_{34} = d\left(  \cos(\theta_1/2),  \sin(\theta_1/2), 0 \right)$. 
%
%

\subsection{Tight-binding model}
\label{sec:tbmodel}

To calculate the electronic structure we consider an effective four-band TB Hamiltonian 
\cite{ruka14,ruyu+15}, containing four $p_z$-like orbitals per unit cell (one per atom), namely
\begin{equation}
    \mathcal{H}_{\rm el} = \sum_{i \neq j} t_{ij} c_{i}^{\dagger} c_{j}\;,
\end{equation}
where indices $i$ and $j$ run over all sites of the phosphorene sheet and $t_{ij}=t_{ji}$ are the hopping parameters. 
Our aim is to develop a simple TB model that accurately describes the strain-induced reconstruction of the electronic 
bands close to the $\Gamma$-point. 

Let us start by putting forward a minimal model that captures the energy dispersion close 
to the $\Gamma$-point in the absence of strain, as schematically shown in Fig. \ref{fig:disp-sketch}(b).
To describe the electron-hole asymmetry and the anisotropy of the electronic dispersion relation around the 
$\Gamma$-point\cite{line+14,qiko+14,feya14,pewe+14}, we consider both nearest and next-nearest neighbor 
couplings within a cutoff radius of $4\, {\rm \AA}$ to obtain the five-parameter TB model put forward in 
Ref.\ \onlinecite{ruka14}. The hopping parameters are denoted by $t_1, \cdots, t_5$. 
Fig.\ \ref{fig:struc}(a) shows the correspondence between the hopping parameters and interatomic matrix elements 
connecting orbitals centered at the $i$ and $j$ lattice atoms. 
We note that $t_5$ does not affect the dispersion of the valence and conduction bands and, hence, set $t_5=0$ to 
simplify the model. 
The agreement between the resulting four-parameter model and DFT band-structure calculations at 
the $\Gamma$-point with respect to the corresponding effective electron and hole masses\cite{line+14} is not very good. 
We find that the accuracy of the TB model is significantly improved by introducing a single additional hopping 
parameter $t_8$, see Fig.\ \ref{fig:struc}(a). 
Finally, we introduce another hopping parameter ($t_6$ in Fig.\ \ref{fig:struc}) to allow for reproducing the strain-dependence of the
electronic band-gap reported in recent \emph{ab initio} calculations \cite{pewe+14}.

At the $\Gamma$-point our six-parameter TB Hamiltonian has the following eigenenergies
\begin{subequations}
\begin{align}
    E_1 ={}&  2 t_1 - t_2 + 2 t_4  - t_6 + 2 t_8\;,\\
    E_2 ={}& -2 t_1 - t_2 - 2 t_4 - t_6 - 2 t_8\;,\\
    E_3 ={}&  2 t_1 + t_2 + 2 t_4 + t_6 + 2 t_8\;,\\
    E_4 ={}& -2 t_1 + t_2 - 2 t_4 + t_6 - 2 t_8\;.
\end{align}
\end{subequations}
(We have subtracted a common energy shift $E_0 = 2t_3$ from all eigenvalues.)
The band gap is given by the difference of the energy of the conduction band ($E_3$) and the valence 
band ($E_2$) 
\begin{equation}
    E_{\rm g}\equiv E_3-E_2 = 4 t_1 + 2 t_2 + 4 t_4 + 2 t_6 + 4 t_8\;,
\end{equation}
whereas for the other two bands we obtain
\begin{equation}
    E_4 + E_1 = 0\;,\quad E_4-E_1 \equiv \Delta E = 4 (t_2 + t_6) - E_{\rm g}\;.
\end{equation}
Hence, using the band-gap and $\Delta E$ as an input one can only determine two hopping parameters.

Additionally, close to the $\Gamma$-point we find that
	\begin{subequations}
		\label{eq:Ek_quad_AC}
		\begin{align}
		E_2(k_x, 0) \approx{}& -\frac{E_{\rm g}}{2}  - \frac{1}{16} \left[ \Delta E + 16 (t_2-6 t_4) - (\Delta E)^{-1}(E_g-8 t_2+ 16 t_4)^2 \right] \left(a_1 k_x\right)^2 \;,\\
		E_3(k_x, 0) \approx{}& +\frac{E_{\rm g}}{2} + \frac{1}{16} \left[ \Delta E + 16 (t_2-6 t_4) - (\Delta E)^{-1} (E_g-8 t_2+ 16 t_4)^2 \right] \left(a_1 k_x\right)^2 \;
		\end{align}
	\end{subequations}
along the armchair direction and
\begin{subequations}
\label{eq:Ek_quad_ZZ}
\begin{align}
    E_2(0, k_y) \approx{}&  -\frac{E_{\rm g}}{2} - \frac{1}{4}\left(t_1 - 4 t_3 + t_4 + 9 t_8 \right) \left(a_2 k_y\right)^2 \;,\\
    E_3(0, k_y) \approx{}&  +\frac{E_{\rm g}}{2}  - \frac{1}{4}\left(t_1 + 4 t_3 + t_4 + 9 t_8 \right) \left(a_2 k_y\right)^2 \;
\end{align}
\end{subequations}
in the zigzag direction.  

Our results are conveniently cast in terms of the energy-band curvatures (or inverse effective masses), namely, 
$c_{i,\alpha}=\partial^2 E_i/\partial k_\alpha^2$ with $i=2,3$ and $\alpha=x,y$.
Using $E_{\rm g}$, $\Delta E$, $c_{2,y}$ and $c_{3,y}$ as input parameters, we obtain the following simple analytical relations
\begin{subequations}
\label{eq:paramrel}
\begin{align}
t_1+t_4 ={}& \frac{9}{64}\left(E_{\rm g}-\Delta E\right)-\frac{1}{8} a_2^{-2} (c_{3,y}+c_{2,y}) \;,\\
t_2+t_6 ={}& \frac{1}{4} \left(E_{\rm g} + \Delta E\right) \;,\\
t_3 ={}&  \frac{1}{4} a_2^{-2}(c_{3,y}-c_{2,y}) \;,\\
t_8 ={}&  \frac{1}{64} \left(\Delta E-E_{\rm g}\right)+\frac{1}{8} a_2^{-2} (c_{3,y}+c_{2,y}) \;.
\end{align}
\end{subequations}
These relations show that one needs more input information to uniquely determine the model 
parameters. We choose to use the curvature in armchair direction and the strain-dependence of 
the band-gap energy, a quantity that we address in the following section.
 
\subsection{Strained phosphorene}
\label{sec:strained}

The application of strain to a phosphorene sheet induces a shift in the interatomic distances, which also changes the material 
band-structure. In the TB model, the modifications of the electronic properties can be accounted for by hopping 
parameters $t_{ij}$ which depend on the interatomic distances. 
Here, we assume that after a deformation the hopping parameters become
\begin{equation}
\label{eq:tstrain}
t'_{ij} = t_{ij} \exp[-\beta (|\vec{R}'_{ij}|/|\vec{R}_{ij}|-1)]\;, 
\end{equation}
where $\vec{R}'_{ij}$ ($\vec{R}_{ij}$) is the modified (original) vector connecting atoms at sites $i$ and $j$ and 
$\beta$ quantifies the decay.

Next, we need to establish a relation between the applied strain, which is a macroscopic 
quantity, and the resulting microscopic shift in atomic positions. As already mentioned, since phosphorene 
has four atoms in the unit cell, the Cauchy-Born rule does not apply\cite{er08,mile+16}. 
Instead, the position 
of the atoms within the unit cells of a strained sample is obtained by minimizing the 
microscopic elastic energy with the constraint $\vec{a}'_i  = \vec{a}_{i} + \ten{\strain}\cdot
\vec{a}_{i}$ where $\vec{a}_i$ is the unstrained primitive lattice vector, $\vec{a}'_i$ is the strained 
one, and $\ten{\strain}$ is the strain tensor. The latter has three independent elements and is given by
\begin{equation}
\ten{\strain} = \left(
\begin{matrix}
    \strain_{xx} & \strain_{xy} & 0 \\
    \strain_{xy} & \strain_{yy} & 0 \\
    0   &   0 & 0 
\end{matrix}
\right)\;.
\end{equation}
These elements lead to the following set of strain-displacement relations for phosphorene\cite{mile+16}
\begin{subequations}
\label{eq:sdr}
\begin{align}
    \vec{b}'_{12} =& \vec{b}_{12} + \ten{\strain}\cdot \vec{b}_{12} - \vec{v}_{||} \;,\\
    \vec{b}'_{23} =& \vec{b}_{23} + \ten{\strain}\cdot \vec{b}_{23} + \vec{v}  \;,\\
    \vec{b}'_{34} =& \vec{b}_{34} + \ten{\strain}\cdot \vec{b}_{34} - \vec{v}_{||}  \;,
\end{align}
\end{subequations}
where $\vec{b}_{ij}$ are the bond vectors connecting the atoms $i$ and $j$ within the unit cell. 
The vector $\vec{v}$ reads
\begin{equation}
\label{eq:v}
\vec{v} = \left(
\begin{matrix}
\kappa_1 u_{xx} + \kappa_2 u_{yy} \\
\kappa_3 u_{xy} \\
\kappa_4 u_{xx} + \kappa_5 u_{yy}
\end{matrix}
\right)\;,
\end{equation}
and $\vec{v}_{||}$ is the projection of $\vec{v}$ onto the plane of the monolayer. The parameters $\kappa_i$ 
are obtained by minimizing the elastic energy per unit cell with the constraints imposed by Eqs.\ \eqref{eq:sdr}. 
Note that the component $v_z$ accounts for a transversal Poisson effect, i.e., a change of thickness due to 
strain in the $x-y$ plane. In phosphorene, this effect has recently been addressed by using an {\it ad hoc} 
``3D Cauchy-Born'' relation, where the strain-tensor has a non-zero component ${u}_{zz}$ \cite{Roldan2015}.  
In effect, this procedure correctly accounts for the component $v_z$, but neglects the components $v_x$ and $v_y$.

To estimate the modification of the hopping terms due to an applied strain, Eq.~\eqref{eq:tstrain},
 it is convenient to write $\vec{R}_{ij}$ 
as a linear combination of the lattice vectors and the bond vectors in the unit cell, namely
\begin{equation}
    \vec{R}_{ij}=n \vec{a}_1+m \vec{a}_2 + \sum_{k} C^k_{ij} \vec{b}_k\;,
\end{equation}
where the index $k$ runs over the pairs of sites $(12),\;(23)$, and $(34)$. $C_{ij}^{k}$ is a matrix of 
integers, which is explicitly given in Appendix \ref{app:SDR}. After a little algebra, we write $|\vec{R}'_{ij}|$ as
\begin{equation}
\label{eq:dlength}
|\vec{R}'_{ij}|=|\vec{R}_{ij}| + \frac{\vec{R}_{ij}\cdot \ten{\strain}\cdot\vec{R}_{ij}}{|\vec{R}_{ij}|} + 
\frac{\left[C^{ij}_{(23)}\vec{v}-(C^{ij}_{(12)}+C^{ij}_{(34)})\vec{v}_{||}\right]\cdot\vec{R}_{ij}}{|\vec{R}_{ij}|}\;.
\end{equation}
We comment at this point on the implications of Eq.\ \eqref{eq:dlength} on the modification of the dispersion 
due to an applied shear. Since the component $v_y$ is proportional to the shear, the last term in 
Eq.\ \eqref{eq:dlength} implies that shear, in contrast to what one expects from the Cauchy-Born rule, 
does not preserve the lengths of bond vectors with a non-vanishing $y$-component. 
On the other hand, the inversion symmetry of phosphorene implies that shear cannot contribute to the 
electronic structure to linear order. Consistent with this, we find that the contributions to the electronic 
structure from an applied shear cancel to linear order, leaving the band structure unchanged for small shear\cite{sali+15}.

The band structure of strained phosphorene is obtained by using the modified hopping terms $t_{ij}'$
from Eq.\ \eqref{eq:tstrain} in the TB Hamiltonian. In particular, the band-gap of strained phosphorene is given by 
\begin{equation}\label{eq:gap_str}
    E_{\rm g}' = 4 t_1' + 2 t_2' + 4 t_4' + 2 t_6' + 4 t_8'\;.
\end{equation}
To lowest order in strain, the modified effective masses are obtained by inserting $t'_{ij}$ 
into Eqs.\ \eqref{eq:Ek_quad_AC} and \eqref{eq:Ek_quad_ZZ}, which describe the electronic dispersion 
close to the $\Gamma$-point.

\section{Results}\label{sec:results}
\subsection{Parameter estimation}

We have described the general procedure for obtaining a TB model to calculate strain-induced 
changes on the electronic structure of phosphorene. Let us now determine the model parameters 
introduced in the previous section.  

We obtain the strain-displacement parameters $\kappa_i$ by minimizing the VFM put forward in 
Ref.\ \onlinecite{micr16a}. 
The results are given in Table \ref{tab:1}. Next, with the help of Eq.\ \eqref{eq:gap_str} we write the 
strain-induced modification of the band-gap, $\Delta E_{\rm g} = E_{\rm g}' - E_{\rm g}$, as
\begin{equation}\label{eq:gap2}
    \Delta E_{\rm g} \approx \beta \strain_{xx} (0.11 t_1 + 0.10 t_2 - 4.63 t_4 - 0.65 t_6 + 0.02 t_8) + 
                             \beta \strain_{yy} (-1.50 t_1 + 0.13 t_2 - 1.64 t_4 + 0.33 t_6 - 3.54 t_8)\;.
\end{equation}

\begin{figure}[t]
        \includegraphics[width=0.33\textwidth]{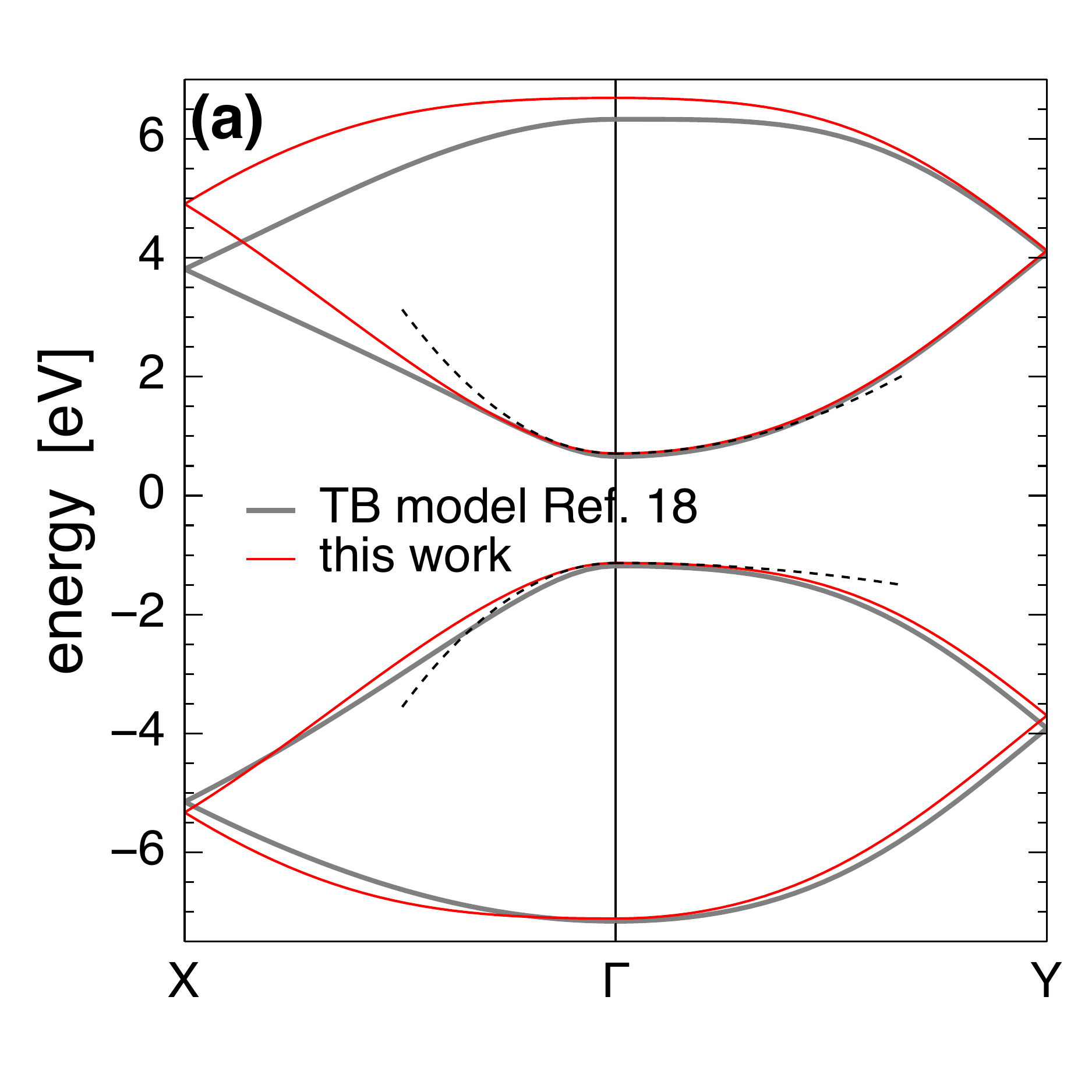}\hfill
        \includegraphics[width=0.3\textwidth]{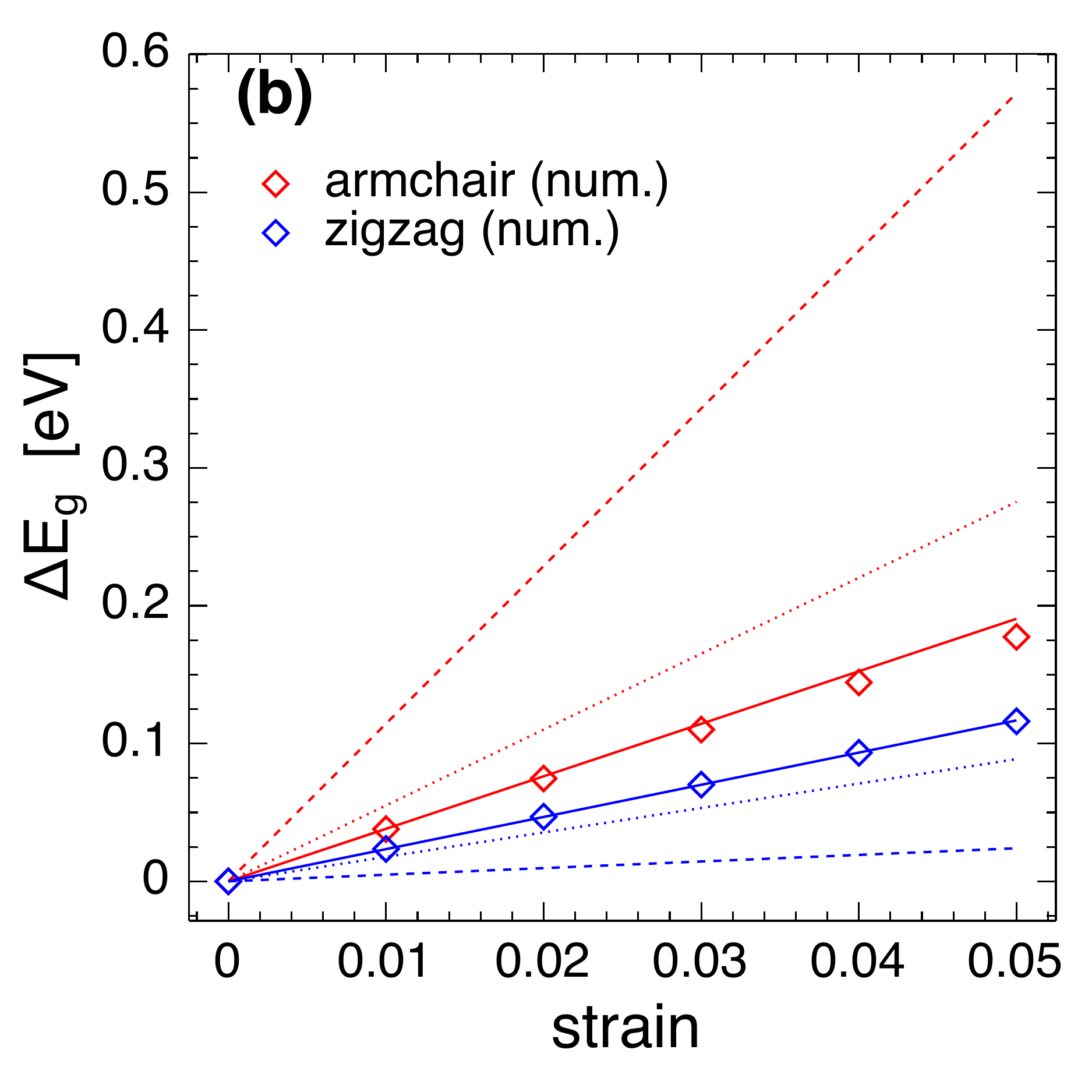}\hfill
        \includegraphics[width=0.3\textwidth]{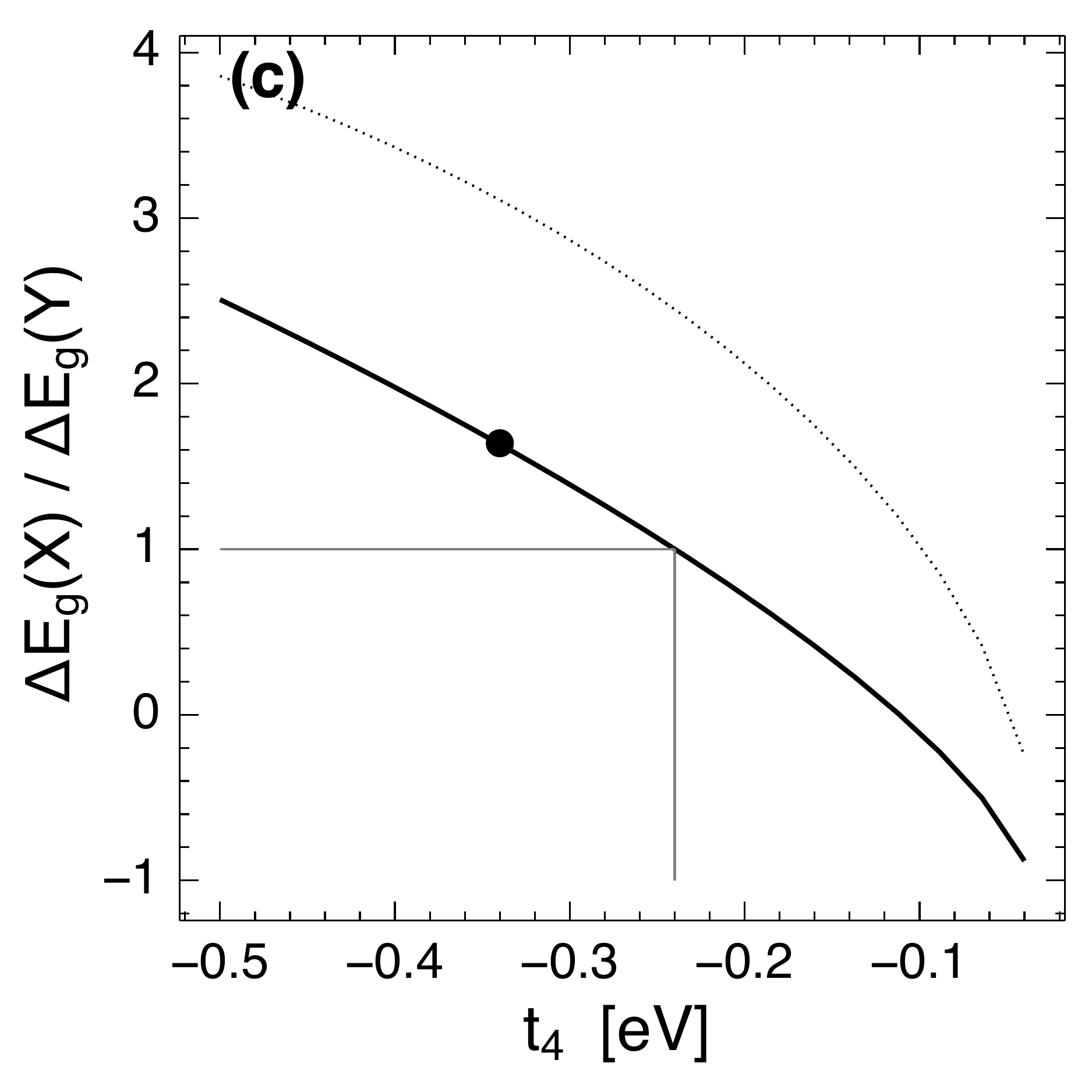}
    \caption{\label{fig:band} (a) Electronic band-structure calculated within the TB model at zero strain. The dashed lines show
        the quadratic dispersions given by Eqs.\ \eqref{eq:Ek_quad_AC} and \eqref{eq:Ek_quad_ZZ}. 
        (b) Strain-induced change of band-gap. Symbols denote fully numerical results, while the solid lines show
        the behavior according to Eq.\ \eqref{eq:gap2} with $\beta=2$ and hopping parameters given in Table \ref{tab:1}.
        The dotted and dashed lines indicate results when using the 2D and 3D Cauchy-Born relations, respectively.
        (c) Ratio of $\Delta E_{\rm g}$ for stretching in armchair and zigzag direction as a function of the hopping 
        parameter $t_4$. A ratio larger than one is only obtained for $t_4 < -0.24\,{\rm eV}$ as shown by the thin lines.}
\end{figure}

The hopping parameters are estimated by fitting the main features of the low energy band structure \cite{ruyu+15},
namely, $E_{\rm g}=1.84\;{\rm eV}$ and the effective masses $m_{\rm c}(X)\approx 0.2 m_{\rm e}$, 
$m_{\rm v}(X)\approx 0.2 m_{\rm e}$, $m_{\rm c}(Y)\approx 1.2 m_{\rm e}$, and $m_{\rm v}(Y)\approx 3.9 m_{\rm e}$. 
Further, we set $E_4=-E_1=6.9\;{\rm eV}$. 
As explained in Sec.\ \ref{sec:tbmodel}, these inputs are not sufficient to determine all six hopping parameters 
and $\beta$. 
To this end, we compare the strain-induced modification of the band-gap predicted by Eq.\ \eqref{eq:gap2} with 
{\it ab initio} results \cite{pewe+14}. The latter show that the band-gap increases linearly by approximately 
$0.1\,{\rm eV}$ with strains up to $\approx 4\%$. The increase is larger for strain in armchair direction. By using 
$\beta=2$ and the hopping parameters given in Table \ref{tab:1} our TB model nicely reproduces this behavior, 
as shown in Fig.\ \ref{fig:band}(b). 
The model is also in good quantitative agreement with the electronic dispersion of Ref.\ \onlinecite{ruyu+15} 
at zero strain, see Fig.\ \ref{fig:band}(a).

We verify the accuracy of our analytical results by comparing the band gaps predicted by Eq.\ \eqref{eq:gap2} with
those obtained from numerical calculations. For that, we consider systems periodic boundary conditions with supercells 
with $15\times 15$ unit cells. For a given strain, the system is allowed to relax using the VFM developed in 
Ref.\ \onlinecite{micr16a}. The band gaps obtained by diagonalization of the TB Hamiltonian for the resulting atomic
configurations agree very well with Eq.\ \eqref{eq:gap2}, as shown by Fig.\ \ref{fig:band}(b). The 2D and 3D 
Cauchy-Born relations (dotted and dashed lines) lead to significant deviations from the numerical results, with the 
latter giving rise to the largest errors.

Interestingly, we find that the magnitude of $t_4$ (or $t_6$ depending on which parameter is undetermined) 
is crucial to correctly describe the behavior of the band gap versus strain. Maintaining the constraints 
for the band-gap and the effective masses, we use Eq.\ \eqref{eq:gap2} to calculate the ratio of $\Delta E_{\rm g}$
for stretching along the armchair and zigzag directions. Figure \ref{fig:band}(c) shows that if  
$t_4 \agt -0.24\,{\rm eV}$, the ratio becomes smaller than one, which disagrees with the findings 
of Ref.\ \onlinecite{pewe+14}. In the present case, $t_4=-0.34\,{\rm eV}$ leads to a ratio of $\approx 1.64$.  
Further investigations and, in particular, experimental results are needed to better assess $\Delta E_{\rm g}$.

\begin{table}[b]
\centering
\begin{tabular}{ c|c|c|c|c|c||c||c|c|c|c|c } 
    $t_1$ (eV) & $t_2$ (eV) & $t_3$ (eV) & $t_4$ (eV) & $t_6$ (eV) & $t_8$ (eV) & $\beta$ &  $\kappa_1$  & $\kappa_2$ & $\kappa_3$ & $\kappa_4$ & $\kappa_5$  \\ 
\hline
-1.25 & 4.38 & -0.106 & -0.34 & -0.47 & 0.09 & 2 &0.71 & 0.27 & 1.26 & -0.39 & -0.16 \\ 
 \hline
\end{tabular}
\caption{Hopping parameters and strain-displacement parameters used in this study}
\label{tab:1}
\end{table}

\subsection{Nonuniform strain}

Now we turn to an example of practical interest: a suspended phosphorene drum and the influence
of nonuniform strain. If the drumhead is subjected to uniform pressure, it will be statically deformed, 
which leads to a nonuniform strain distribution. This system is conveniently described by continuum 
elasticity\cite{micr16a}. The phosphorene monolayer is effectively modeled as a thin anisotropic 
plate. Its macroscopic elastic properties are characterized by two bending rigidities and four stiffness 
constants. 
We find the shape of the deformed drum and the strain distribution by solving the equations corresponding 
to the out-of-plane displacement-field $w(\vec{r})$ and for the Airy stress function $\chi(\vec{r})$
\cite{micr16a}.
In following we focus our discussion on approximate analytic solutions (details are given in 
Appendix \ref{app:DrumStrain}) and compare them with numerical results.

For sufficiently small pressure the bending contribution to the elastic energy of the drum dominates and
its shape is well approximated by $w(x,y) = w_0 [1-(x/R)^2-(y/R)^2]^2$, where $w_0$ is the deflection at 
the center and $R$ is the radius of the drum. In contrast, for large pressures stretching plays a major role 
and the shape is approximately  given by $w(x,y) = w_0 [1-(x/R)^2-(y/R)^2]$. 
As shown in Appendix \ref{app:DrumStrain}, the
strain distribution in both regimes becomes
\begin{subequations}
\begin{align}
    \strain_{xx}(\vec{r}) &= (w_0/R)^2\left[\frac{P_1(\vec{r})}{C_{11}}-\frac{C_{12}}{C_{11}C_{22}} P_2(\vec{r})\right] \;,\\
    \strain_{yy}(\vec{r}) &= (w_0/R)^2\left[\frac{P_2(\vec{r})}{C_{22}}-\frac{C_{12}}{C_{11}C_{22}} P_1(\vec{r})\right] \;,\\
    \strain_{xy}(\vec{r}) &= (w_0/R)^2\frac{x y}{R^2}  \frac{P_3(\vec{r})}{C_{66}} \;,
\end{align}
\end{subequations}
where $P_1$, $P_2$ and $P_3$ are polynomials containing even powers of $x/R$ and $y/R$ 
(see Appendix \ref{app:DrumStrain}). The tensile strains are maximal and the shear vanishes at 
the center of the drum. 
From Eq.\ \eqref{eq:gap2} we find that the corresponding change in band-gap is maximal and proportional 
to $(w_0/R)^2$, with a factor of proportionality of approximately $6.4\,{\rm eV}$ for both deformation regimes. 

\begin{figure}[t]
    \includegraphics[width=0.9\textwidth]{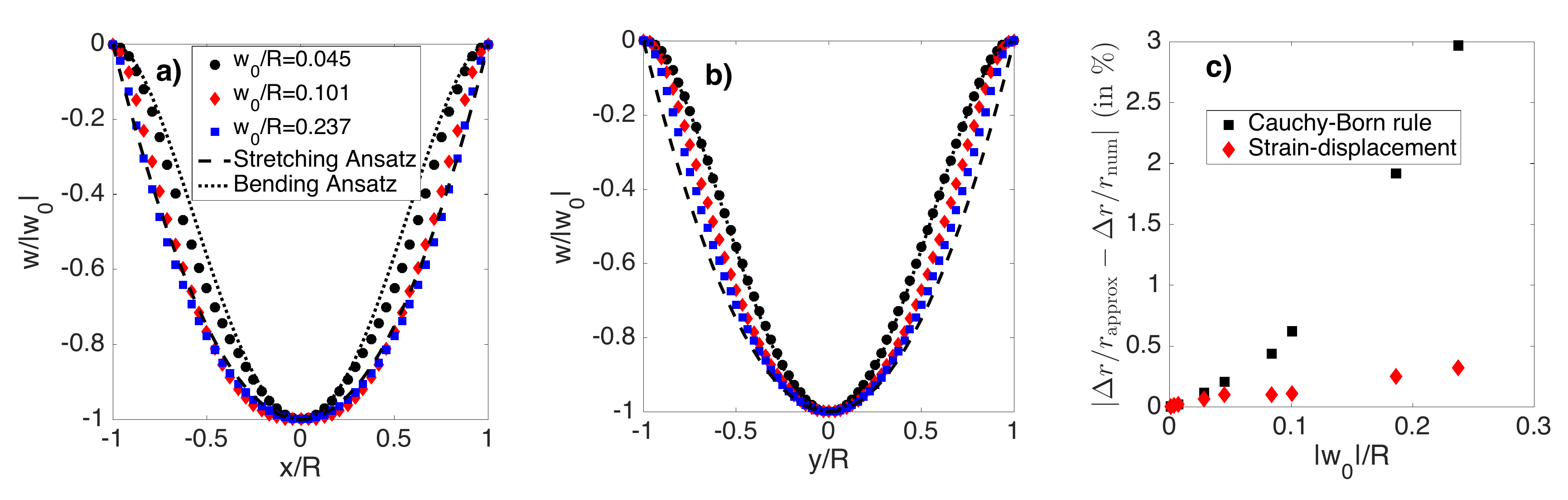}

    \caption{\label{fig:def} a) and b): Deflection of the drum along the lines $y=0$ and $x=0$ for central deflections $w_0/R=0.045$ (black circles), $0.101$ (red diamonds) and $0.237$ (blue squares). The dashed and dotted lines correspond to the stretching and bending Ansatz, respectively. c) The relative error of the estimated bond length at the center of the drum using the 3D Cauchy-Born rule (black squares) and the strain-displacement relations according to Eq.\ \eqref{eq:sdr} (red diamonds).}
\end{figure}

In order to verify the analytical results, we use the VFM of Ref.\ \onlinecite{micr16a} to numerically calculate 
the shape of a pressurized drum. To validate our approach in the continuum limit while maintaining computational
convenience we simulate a drum of radius $103.16$ \AA{}. 
The out-of-plane deformation field $w(\vec{r})$ is obtained by considering the midpoints of each primitive 
unit cell. In Figs.\ \ref{fig:def}(a) and \ref{fig:def}(b) the deformation field is plotted along $x=0$ and $y=0$, respectively,
for three central deflections $w_0/R$, together with the analytical expressions for the deformation in the bending and 
stretching regimes. We find that the drum shape is indeed very close to being radially symmetric and shows 
the expected crossover from the bending to the stretching dominated regime as $w_0/R$ increases.

Further, we consider the vectors connecting the midpoints of adjacent primitive unit cells to estimate the local strain 
distribution. We test the accuracy of our strain-displacement relations,
by calculating the expected lengths of the bond vectors using Eq.\ \eqref{eq:sdr}, and comparing them to the 
numerically obtained ones. The maximal error of the strain-displacement relations occurs at the center of the drum, 
where the strain is maximal.  Fig.\ \ref{fig:def}(c) shows the relative errors as a function of central deflection 
together with the relative error given by the 3D Cauchy-Born rule, ignoring $\vec{v}_{||}$. At the largest deflection, 
the accuracy of the Cauchy-Born approximation is about $3\%$ at the largest deflection, whereas 
our method leads to bond length relative errors of less than $0.5\%$. These differences have a significant impact 
on the hopping matrix elements $t_{ij}$ and, hence, on the tight-binding electronic structure calculations. 
%

\begin{figure}[t]
    \includegraphics[width=0.45\textwidth]{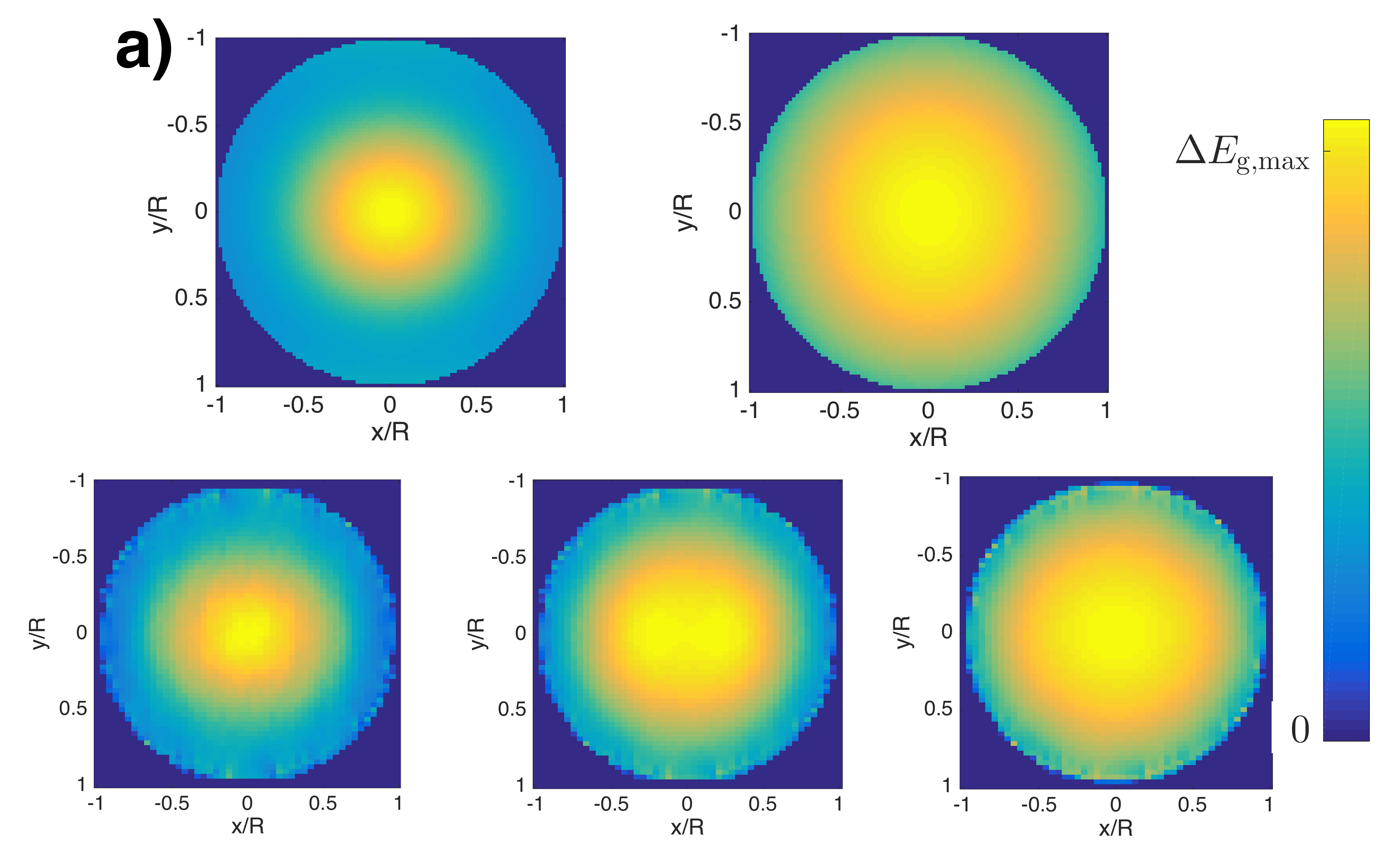}
    \includegraphics[width=0.45\textwidth]{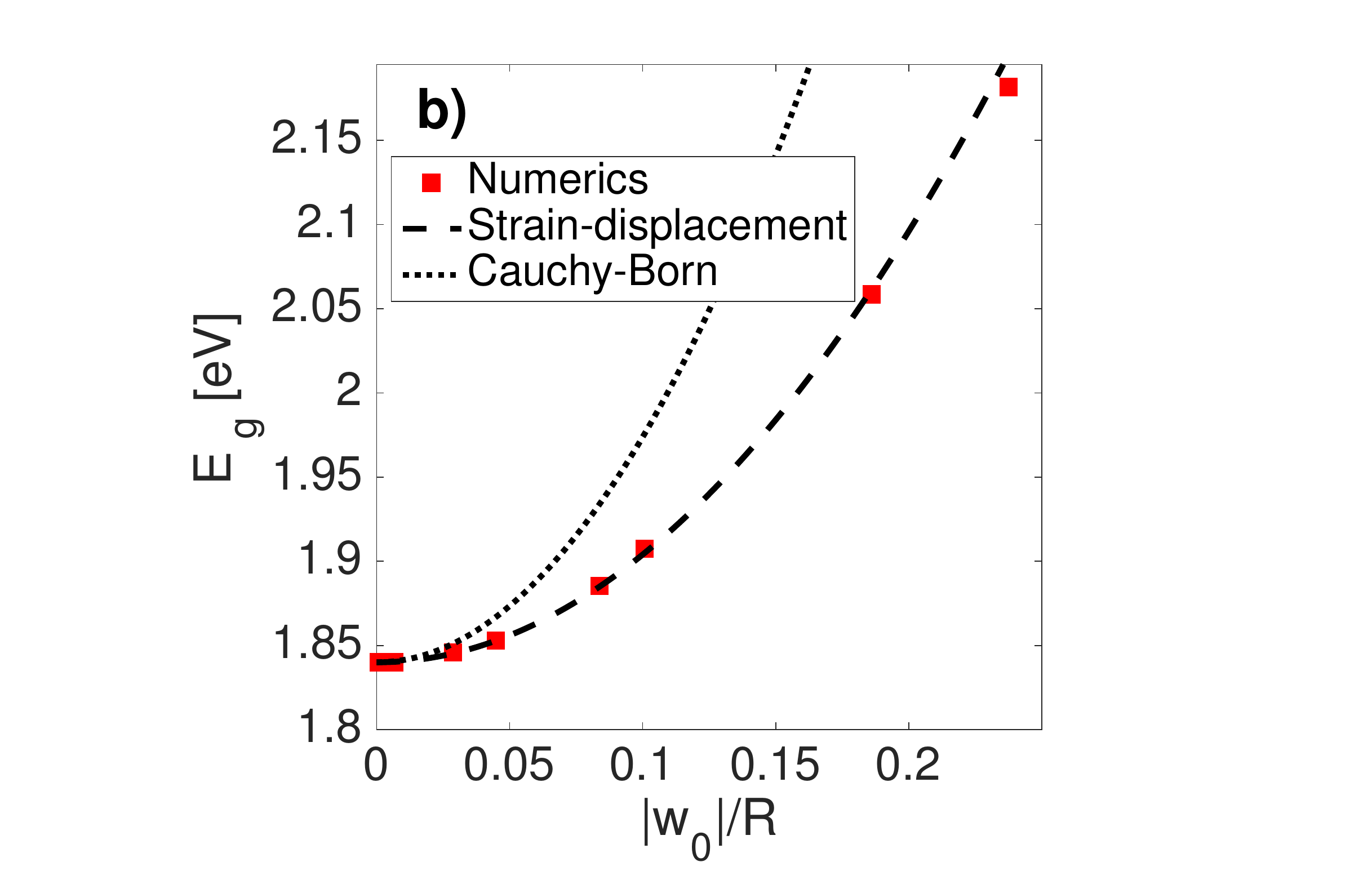}
    \caption{\label{fig:loc_gap} Local band-gap of the pressurized drum. 
    a) The three colormaps in the bottom row show numerically obtained
        results for different $w_0/R=0.045,0.101,0.237$.
        The colormaps in the upper row are obtained with the bending and the stretching Ansatz, respectively.
    b) Numerical band-gap at the center of the drum (red squares) and according to Eq.\ \eqref{eq:gap2} (dashed line).}
\end{figure}

For each unit cell we also calculate the (local) electronic band-structure using the TB model and a
$3\times 3$ super-cell centered at the unit cell of interest and using periodic boundary conditions.
This allows us to obtain the local band-gap $E_{\rm g}(\vec{r})$. In Fig.\ \ref{fig:loc_gap}(a) we show maps
of the band-gap for three different central deflections, indicating the transition from a bending-dominated 
regime at low deflections to a stretching dominated regime at high deflections. Figure \ref{fig:loc_gap}(b)
shows a comparison between the numerically calculated band-gap at the center of the drum with our 
analytical result. The agreement is excellent, suggesting that the model can be used to give qualitative 
predictions of local electronic properties in complex geometries. 

\section{Conclusions}
\label{sec:conclusions}

In summary, we have presented a multi-scale approach to calculate electronic properties of deformed 
phosphorene. To this end, we developed a TB model to describe the electronic structure. We found
that six hopping parameters are required to get a good quantitative description of the electronic bands, 
which includes the band-gap and the (anisotropic) effective masses at the $\Gamma$-point. The influence 
of deformations is described by considering distance-dependent hopping-parameters. 
The crucial point of our approach is the fact that in almost all cases of interest, the macroscopic strain-distribution 
can be inferred, but the microscopic positions of the atoms after deformation are not known.
Thus a relation between strain and atomic displacements is required to make use of the TB model. 
As we have shown, the simple Cauchy-Born relation - often employed in this context - is not a good 
approximation as it tends to overestimate the strain. 
Instead, we propose to use strain-displacement relations, which are obtained by minimizing the elastic 
energy\cite{mile+16}. This approach is easily combined with the microscopic TB model and 
yields good quantitative agreement. The central result is given by Eq.\ \eqref{eq:gap2}, which gives 
an expression for the strain-induced modification of the band-gap for a given (homogeneous) strain.

We demonstrate our method for the relevant case of a phosphorene drum. The deformation due to 
pressure leads to inhomogeneous strain distributions. Introducing a local band-gap, one then finds a 
spatially changing strain-induced contribution, which is essential for the so-called inverse funneling 
effect\cite{sapa+16}.
We obtain a very good agreement between fully numerical results (TB and VFM) with analytical estimates 
resulting from continuum elasticity\cite{micr16a} combined with the derived strain-induced modification of the 
band-gap given by Eq.\ \eqref{eq:gap2}.

The multi-scale approach presented in this article can also be used for other 2D materials, provided 
suitable TB and VF models are known. It offers a quantitative and efficient procedure to study the 
structural and electronic properties of single-layered materials with arbitrary deformation landscapes.

\begin{acknowledgements}
    C.H.L.\ acknowledges financial support of the Brazilian funding agencies CNPq and FAPERJ.
\end{acknowledgements}

\appendix
\section{Strain-displacement relation for phosphorene}
\label{app:SDR}
%
A detailed description of the procedure to obtain strain-displacement relations,
Eqs.~\eqref{eq:sdr}, from a VFM is given in Ref.\ \onlinecite{mile+16}. Here we revisit the theory 
to explain and provide further insight on some of the expressions used in Sec.\ \ref{sec:strained}.

\begin{figure}[t]
    \includegraphics[width=0.45\textwidth]{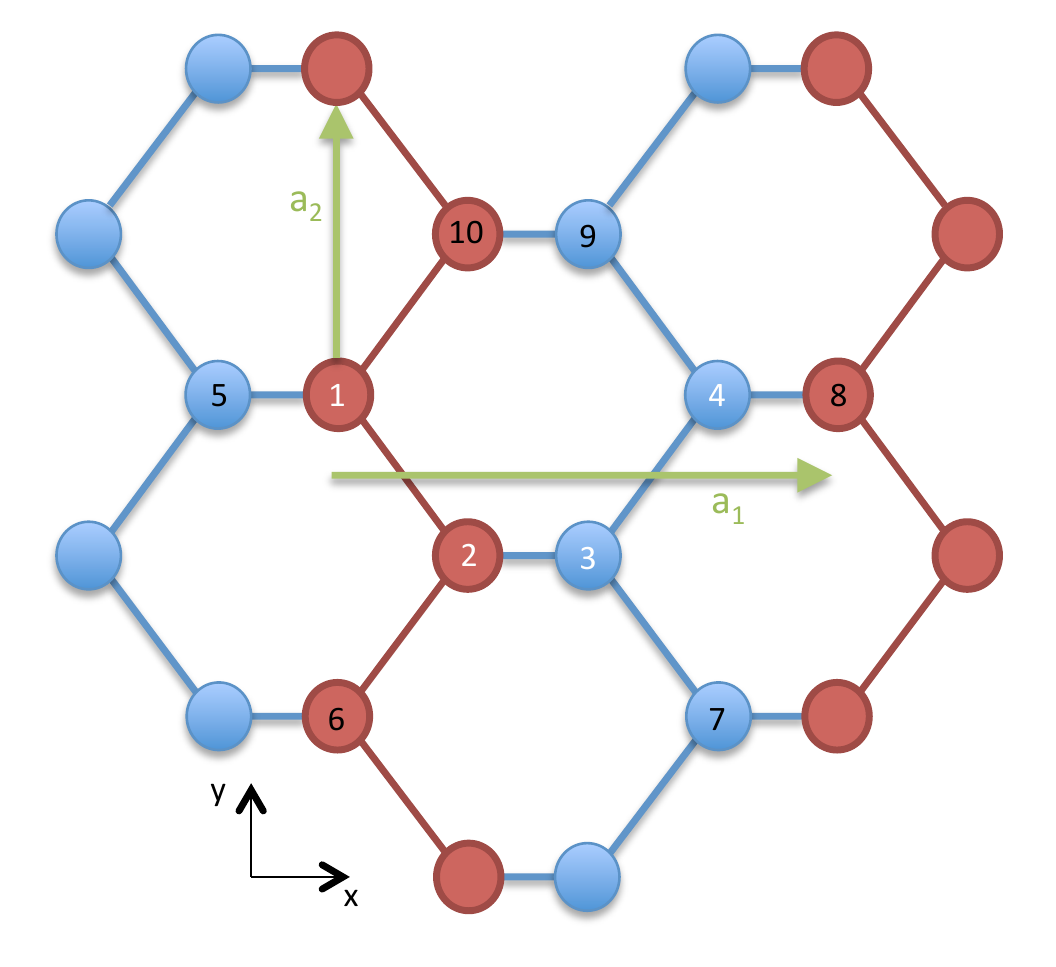}
    \caption{\label{fig:struc-vecs} Sketch of phosphorene structure indicating the nearest neighbors ($5$ to $10$) of the atoms in the unit cell ($1$ to $4$).}
\end{figure}

For a homogeneous strain the primitive lattice-vectors change according to
\begin{equation}\label{eq:trans_prim_vecs}
    \vec{a}_i' = \vec{a}_i + \ten{u}\cdot\vec{a}_i\;.
\end{equation}
The position $\vec{R}_i$ of any arbitrary lattice site can be expressed in terms of the
primitive lattice-vectors and the $N_{\rm bb}$ bond-vectors of the basis,
\begin{equation}
    \vec{R}_i = m_i \vec{a}_1 + n_i \vec{a}_2 + \sum^{N_{\rm bb}}_{k=1} C_{i}^k \vec{b}_k\;,
\end{equation}
where $m_i$, $n_i$, and the entries of the matrix $C_{i}^k$ are integers. 
The vector $\vec{R}_{ij}$ connecting the atoms $i$ and $j$ is then given by
\begin{equation}
    \vec{R}_{ij} = m_{ij} \vec{a}_1 + n_{ij} \vec{a}_2 + \sum^{N_{\rm bb}}_{k=1} C_{ij}^k \vec{b}_k\;,
\end{equation}
where for all variables $X_{ij}=X_i-X_j$ holds. Since we assume that the deformation preserves the 
lattice structure, $m_{ij}$, $n_{ij}$ and $C_{ij}^k$ remain unchanged, upon strain 
\begin{equation}\label{eq:trans_vecs}
    \vec{R}_{ij}' = \vec{R}_{ij} + \ten{u}\cdot\vec{R}_{ij}  + \sum^{N_{\rm bb}}_{k=1} C_{ij}^k \vec{\delta b}_k\;,
\end{equation}
which implies that the basis bond-vectors transform as
\begin{equation}
    \vec{b}_k' = \vec{b}_k + \ten{u}\cdot\vec{b}_k + \vec{\delta b}_k\;.
\end{equation}
For a monoatomic basis, where $N_{\rm bb}=0$, Eq.\ \eqref{eq:trans_vecs} is reduced to the standard Cauchy-Born 
relation. 
The lattice symmetries relate the vectors $\vec{\delta b}_k$ to each other and thus reduce the number of unknown 
components\cite{mile+16}.
In the case of phosphorene, one has $\vec{\delta b}_{23}=\vec{v}$ and 
$\vec{\delta b}_{12}=\vec{\delta b}_{34}=\vec{v}_\parallel$, which are explicitly given by Eq.~\eqref{eq:v}.
The nearest neighbors of the atoms within a unit cell are shown in Fig.\ \ref{fig:struc-vecs}. The corresponding values 
of $m_{ij}$, $n_{ij}$ and $C_{ij}^k$ are given in Table \ref{tab:2}.

In general, the vectors $\vec{\delta b}_k$ can be determined by minimizing the elastic energy of the deformed structure,
that for small deformations reads\cite{asme76}
\begin{equation}
    E_{\rm VFM} = \frac{1}{2}\sum_{ij,i'j'} \left(\vec{R}_{ij}'-\vec{R}_{ij}\right) \cdot \ten{K}_{ij,i'j'} \cdot\left(\vec{R}_{i'j'}'-\vec{R}_{i'j'} \right)\;,
\end{equation}
where the sum runs over all pairs of atoms and $\ten{K}_{ij,i'j'}$ denotes the matrix of force-field parameters. 
The latter is typically restricted to neighboring bonds. Using Eq.\ \eqref{eq:trans_vecs} one finds that $E_{\rm VFM}$ 
is a second order polynomial in $\ten{u}$ and $\vec{\delta b}_k$. 
By requiring that $\partial E_{\rm VFM}/\partial \vec{\delta b}_k =0$ one obtains a set of linear relations between 
$\vec{\delta b}_k$ and $\ten{u}$, which constitutes the strain-displacement relations.

\begin{table}
    \centering
    \begin{tabular}{c|cc|ccc}
        $(i,j)$ & $n_{ij}$ & $m_{ij}$ & $C_{ij}^{(12)}$  & $C_{ij}^{(23)}$  & $C_{ij}^{(34)}$ \\
        \hline
        $(1,2)$ & $0$ & $0$ & $1$ & $0$ & $0$\\
        $(2,3)$ & $0$ & $0$ & $0$ & $1$ & $0$\\
        $(3,4)$ & $0$ & $0$ & $0$ & $0$ & $1$\\
        \hline
        $(1,5)$ & $-1$ & $0$ & $1$ & $1$ & $1$\\
        $(2,6)$ & $0$  & $-1$ & $-1$ & $0$ & $0$\\
        $(3,7)$ & $0$  & $-1$ & $0$ & $0$ & $1$\\
        $(4,8)$ & $-1$ & $0$ & $1$ & $1$ & $1$\\
        $(4,9)$ & $ 1$ & $0$ & $-1$ & $-1$ & $-1$\\
        $(1,10)$ & $0$ & $1$ & $1$ & $0$ & $0$\\
        \hline
    \end{tabular}
    \caption{Decomposition of the nine nearest-neighbor vectors within the unit cell and connecting to the neighboring unit cells
        according to Eq.\ \eqref{eq:trans_vecs}.}
    \label{tab:2}
\end{table}

\section{Strain distribution for a drum}
\label{app:DrumStrain}
Denoting the displacement field by $\vec{u}=(u(\vec{r}),v(\vec{r}),w(\vec{r}))$ and the Airy stress function by $\chi(\vec{r})$, the
shape of the deformed drum is determined by the following equations\cite{micr16a}
\begin{align}
\label{eq:airyeq}
    \frac{1}{Y_y}& \partial^4_x \chi + \frac{1}{Y_x} \partial^4_y \chi 
    + \left(\frac{1}{G_{xy}} -2\sqrt{\frac{\nu_{xy}\nu_{yx}}{Y_{y}Y_{x}}} \right) \partial^2_x\partial^2_y \chi=
     (\partial_x \partial_y w)^2-(\partial^2_x w) (\partial^2_y w)\;
\end{align}
and
\begin{align}
\label{eq:w}
     (\partial^2_x \chi) (\partial^2_y w) -2 (\partial_x\partial_y \chi) (\partial_x\partial_y w) =  \left(\kappa_x \partial^4_x w + \kappa_y \partial^4_y w + 2 \sqrt{\kappa_x \kappa_y} \partial^2_x \partial^2_y w\right) - P_z\;.
\end{align}
Here, $Y_{x,y}$ and $\kappa_{x,y}$ are the Young's modulus and the bending rigidity in $x$ and $y$ direction, 
$G_{xy}$ is the shear modulus and $\nu_{xy}$ and $\nu_{yx}$ are the Poisson ratios.

In the following, we solve these equations for a drum of radius $R$ with vanishing prestrain under a spatially 
constant external pressure for the limit cases of small and large deformations. Taking into account the 
boundary conditions for the in-plane displacement fields, namely, $u(\tilde{x}=0,\tilde{y}=0)=0$, 
$v(\tilde{x}=0,\tilde{y}=0)=0$, $u(\tilde{x}^2+\tilde{y}^2=1)=0$, $v(\tilde{x}^2+\tilde{y}^2=1)=0$, we 
find the Airy stress function and, hence, the strain fields.
 
For sufficiently small deformations one can ignore the left hand side in \eqref{eq:w} which is cubic in the deformation. 
The deformation is then given by $w(x,y) = w_0 (1-\tilde{x}^2-\tilde{y}^2)^2$ where $\tilde{x}=x/R$ and $\tilde{y}=y/R$. 
The maximal deflection $w_0$ is related to the applied pressure by
\begin{equation}
\frac{w_0}{R} = \frac{P_z R^3}{64 \kappa_{\rm eff}}\;,
\end{equation}
where $\kappa_{\rm eff} \equiv (3\kappa_x+3\kappa_y + 2\sqrt{\kappa_x\kappa_y})/8\approx 3.9$ eV for
phosphorene\cite{micr16a}.

We consider a generic 8th order polynomial as Ansatz for the Airy stress function and solve Eq.\ 
\eqref{eq:airyeq} together with the boundary conditions for the drum in-plane displacement fields 
by coefficient matching. 
The resulting strain distributions are sixth order polynomials in $\tilde{x}$ and $\tilde{y}$ with coefficients 
that depend on intricate combinations of the elastic constants. 
Taking the explicit values for the elastic constants given in Ref.\ \onlinecite{micr16a}, we find 
\begin{align}
\label{eq:strainbend}
u_{xx}= \frac{w_0^2}{R^2}&\left[ 0.73 + 0.12 \tilde{x}^6 - 3.18 \tilde{y}^2 + 3.69 \tilde{y}^4 - 1.24 \tilde{y}^6 + 
\tilde{x}^4 (-0.07 - 1.99 \tilde{y}^2) +\right. \nonumber\\
&\left.\tilde{x}^2 (-0.37 + 4.94 \tilde{y}^2 - 3.14 \tilde{y}^4))\right]\nonumber\\
u_{yy} = \frac{w_0^2}{R^2}&\left[0.63 - 0.84 \tilde{x}^6 + 0.04 \tilde{y}^2 - 0.22 \tilde{y}^4 + 
0.09 \tilde{y}^6 + \tilde{x}^4 (2.35 - 0.80 \tilde{y}^2) + \right. \nonumber \\
& \left.\tilde{x}^2 (-2.14 + 0.88 \tilde{y}^2 - 0.11 \tilde{y}^4)\right]\nonumber \\
u_{xy}= \frac{w_0^2}{R^2}&\left[\tilde{x} \tilde{y} (2.68 + 1.88 \tilde{x}^4 - 2.99 \tilde{y}^2 + 1.06 \tilde{y}^4 + 
\tilde{x}^2 (-4.32 + 2.71 \tilde{y}^2))\right]\;.
\end{align}

To gain insight into the interpretation of these expressions, we solve the Airy stress function for an isotropic 
material with hexagonal symmetry and small Poisson ratio ($\nu\ll 1$), such as graphene. In that case, we 
find a simpler expression for the strain fields that is independent of the elastic constants, namely
 \begin{align}
 \label{eq:straingrap}
 \strain_{xx} =&\,  \frac{w_0^2}{6R^2} \left [5 - 6 (\tilde{x}^2 + 3 \tilde{y}^2) + 
   4 (\tilde{x}^2 + \tilde{y}^2) (\tilde{x}^2 + 5 \tilde{y}^2) - (\tilde{x}^2 + \tilde{y}^2)^2 (\tilde{x}^2 + 7 \tilde{y}^2)\right]\;,\nonumber \\
 \strain_{yy} = &\,  \frac{w_0^2}{6R^2} \left [5 - 6 (\tilde{y}^2 + 3 \tilde{x}^2) + 
   4 (\tilde{y}^2 + \tilde{x}^2) (\tilde{y}^2 + 5 \tilde{x}^2) - (\tilde{y}^2 + \tilde{x}^2)^2 (\tilde{y}^2 + 7 \tilde{x}^2)\right]\;,\nonumber \\
 \strain_{xy} = &\, \frac{\tilde{x} \tilde{y}}{3}  \left[6 - 8 (\tilde{x}^2 + \tilde{y}^2) + 3 (\tilde{x}^2 + \tilde{y}^2)^2\right]\;.  
\end{align}
Comparing the leading order coefficients of the polynomials in Eqs.\ \eqref{eq:straingrap} and \eqref{eq:strainbend} we find 
that despite the anisotropy of phosphorene, the strain distribution close to the center of the drum, $\tilde{x} \ll 1$ and
$\tilde{y}\ll 1$, deviates only by 
about $30\%$ from what one would expect for an isotropic material.

In the opposite limit of large deformations, we ignore the bending related terms in \eqref{eq:w} and 
consider the Ansatz $w(x,y) =  w_0 (1-\tilde{x}^2-\tilde{y}^2)$, $\tilde{x}=x/R$ and $\tilde{y}=y/R$ for the drum 
deformation. Inserting this Ansatz into \eqref{eq:airyeq} we find that the right-hand-side of the equation becomes 
spatially constant, namely, $(\partial_x \partial_y w)^2-(\partial^2_x w) (\partial^2_y w) = -(4w_0^2/R^4)$. 
By inspection of \eqref{eq:airyeq}, this implies that the Airy function is given by a fourth order polynomial whose
coefficients are obtained by matching the boundary conditions 
at the rim of the drum. We find the strain distribution
\begin{align}
    \strain_{xx} = &\, (2/3-\gamma_x) (w_0/R)^2 (1-\tilde{y}^2) + \gamma_x (w_0/R)^2 \tilde{x}^2\;, \\
    \strain_{xy} = &\, (2/3+\gamma_x+\gamma_y) (w_0/R)^2 \tilde{x} \tilde{y}\;,  \\
    \strain_{yy} = &\, (2/3-\gamma_y)(w_0/R)^2(1-\tilde{x}^2) + \gamma_y (w_0/R)^2 \tilde{y}^2\;,
\end{align}
where $\gamma_x$ and $\gamma_y$ are functions of the elastic parameters. 
For phosphorene one has $\nu_{xy} \ll 1$, for which one finds $\gamma_y\approx 0$ and 
$\gamma_x\approx (2/3)\nu_{yx} \left(C_{66}-2C_{12}\right)/\left(6 C_{12} + C_{66} \nu_{yx}\right)$. 
Using the values for the elastic constants provided in Ref.\ \onlinecite{micr16a} we find that $C_{66}\approx 2C_{12}$, 
and therefore $\gamma_x\approx 0$. Interestingly, despite phosphorene being anisotropic, this result coincides 
with the case of isotropic materials with small Poisson ratio ($\nu\ll 1$), where $\gamma_x=\gamma_y=0$. 

\bibliographystyle{prsty}
\bibliography{BP}

\end{document}